\documentclass[aps,prd,onecolumn,superscriptaddress,nofootinbib,11pt]{revtex4}
\usepackage{epsfig,amsfonts,amsthm}
\usepackage{amsmath,amssymb}
\usepackage[usenames,dvipsnames]{xcolor} 
\usepackage{hyperref}
\hypersetup{
   colorlinks=true,       
    linkcolor=Blue,          
    citecolor=Plum,        
    filecolor=magenta,      
    urlcolor=YellowOrange           
}
\usepackage[utf8]{inputenc}

\newcommand{\be}{\begin{equation}}
\newcommand{\ee}{\end{equation}}
\newcommand{\bea}{\begin{eqnarray}}
\newcommand{\eea}{\end{eqnarray}}
\newcommand{\dst}{\displaystyle}
\newcommand{\fr}[2]{\frac{{\dst #1}}{{\dst #2}}}

\newcommand{\Tr}{\mathrm{Tr}}
\newcommand{\Z}{\mathbb{Z}}
\newcommand{\RR}{\mathbb{R}}

\newcommand{\mmatrix}[4]{ \left(\! \begin{array}{ccc}#1 & #2 \\ #3 & #4 \end{array}\!\right) }
\newcommand{\mmmatrix}[9]{ \left(\! \begin{array}{ccc}#1 & #2 & #3\\ #4 & #5 & #6\\ #7 & #8 & #9\\ \end{array}\!\right) }
\providecommand{\id}{{\boldsymbol{1}}}
\providecommand{\mtrx}[1]{\begin{pmatrix} #1 \end{pmatrix}}

\newcommand{\toCP}{\xrightarrow{CP}}
\definecolor{darkgreen}{rgb}{0.0, 0.6, 0.2}
\definecolor{DARKGREEN}{rgb}{0.0, 0.6, 0.2}

\def\lsim{\mathrel{\rlap{\lower4pt\hbox{\hskip1pt$\sim$}}
    \raise1pt\hbox{$<$}}}         
\def\gsim{\mathrel{\rlap{\lower4pt\hbox{\hskip1pt$\sim$}}
    \raise1pt\hbox{$>$}}}         

\begin{document}
\title{
{\normalsize \hfill CFTP/19-001} \\*[7mm]
Beyond basis invariants}

\author{Igor~P.~Ivanov}\thanks{E-mail: igor.ivanov@tecnico.ulisboa.pt}
\affiliation{CFTP, Instituto Superior T\'{e}cnico, Universidade de Lisboa,
Avenida Rovisco Pais 1, 1049 Lisboa, Portugal}
\author{Celso~C.~Nishi}\thanks{E-mail: celso.nishi@ufabc.edu.br}
\affiliation{Centro de Matem\'atica, Computa\c c\~ao e Cogni\c c\~ao,
Universidade Federal do ABC - UFABC,
09.210-170, Santo Andr\'e, SP, Brazil
}
\author{Andreas~Trautner}\thanks{E-mail: trautner@mpi-hd.mpg.de}
\affiliation{Max-Planck-Institut f\"ur Kernphysik, Saupfercheckweg 1, 69117 Heidelberg, Germany}

\begin{abstract}
Physical observables cannot depend on the basis one chooses to describe fields.
Therefore, all physically relevant properties of a model are, in principle, expressible in terms of 
basis-invariant combinations of the parameters.
However, in many cases it becomes prohibitively difficult
to establish key physical features exclusively in terms of basis invariants.
Here, we advocate an alternative route in such cases:
the formulation of basis-invariant statements in terms of basis-covariant objects.
We give several examples where the basis-covariant path 
is superior to the traditional approach in terms of basis invariants.
In particular, this includes the formulation of necessary and sufficient basis-invariant conditions 
for various physically distinct forms of $CP$ conservation in two- and three-Higgs-doublet models. 
\end{abstract}

\maketitle
\enlargethispage{1cm}

\section{Introduction}

When describing the Standard Model (SM) or
building models beyond the SM, one always faces the notorious freedom of basis-choices.
Complex fields can be rephased, yet the physics emerging from the model 
must be invariant under these rephasings.
In models with several fields with identical quantum numbers, the freedom of basis choices 
is even larger and includes arbitrary rotations in the space of these fields.
One may fix a basis for the initial fields, then arrive at the physical (mass eigenstate) fields
and explore their phenomenology. Or one can switch to a different basis and explore the phenomenology there.
Although the Lagrangian and the intermediate calculations may look vastly different in different bases, 
the observables must be the same.

This utterly obvious statement may look less obvious when one actually gets down
to practical calculations. Parameters of the Lagrangian depend on the basis choice,
and attributing physical importance to them can only be done when the basis choice 
is also specified. For example, in the two-Higgs-doublet model (2HDM) 
\cite{Lee:1973iz,Branco:2011iw}, the two doublets can acquire non-zero vacuum expectation values 
$v_1$, $v_2$, whose ratio is customarily denoted as $\tan\beta = v_2/v_1$.
Although the vast majority of papers on 2HDM phenomenology 
describe measurable quantities in terms of $\tan\beta$,
this parameter is by itself basis-dependent and not an observable.
It can become an observable in bases fixed by additional requirements, 
such as when the $\Z_2$ symmetry responsible for natural flavor conservation is manifest~\cite{Haber:2006ue}.

Another vivid example, still within the class of $N$-Higgs-doublet models (NHDM), 
are the conditions for explicit $CP$-conservation \cite{book}. 
A common approach is to simply define a $CP$ transformation in the space
of complex scalar fields $\phi_a$ $(a = 1, \dots, N)$, 
via $\phi_a(\vec{r}, t) \to \phi_a^*(-\vec{r}, t)$.
Using such a definition, one would naively think 
that the model explicitly violates $CP$ symmetry, 
if the scalar potential contains complex coefficients.
This is, however, not true in general. This standard definition of how $CP$ acts on 
scalar fields is basis-dependent.
One can define the general $CP$ transformation via 
\cite{Ecker:1981wv, Ecker:1983hz, Neufeld:1987wa, Ecker:1987qp}
\be
\phi_a(\vec{r}, t) \to X_{ab} \phi_b^*(-\vec{r}, t)\,,\quad X \in U(N)\,.\label{GCP}
\ee
If a model is invariant under such a transformation
with any matrix $X$, then it is explicitly $CP$ conserving, regardless of whether
the potential has complex coefficients \cite{Grimus:1989qn}.
Although the matrix $X$ does depend on the basis choice,
the \textit{presence} of such a symmetry certainly is a basis-independent fact
and has observable consequences.

These and other simple examples have led the model building community to appreciate 
basis-invariant combinations of the parameters of the Lagrangian, or more simply \textit{basis invariants}. 
The general procedure for construction of such quantities was presented in \cite{Botella:1994cs}:
recognize transformation properties of the parameters under general basis changes,
rewrite them as tensors, and fully contract various tensors to obtain
basis invariants of the model.
A nice illustration of this strategy is given by the NHDM scalar sector
and, in particular, by the issue of $CP$ conservation in the 2HDM
\cite{Botella:1994cs,book,Branco:2005em,Davidson:2005cw,Gunion:2005ja,Varzielas:2016zjc,Trautner:2018ipq}.

There is no doubt that, in any model, all physical observables must be expressible in terms of basis invariants.
A major problem is that, beyond the simplest cases, these expressions become exceedingly
or even prohibitively complicated.
The key message of our paper is: \textit{it is not obligatory to formulate physically relevant,
basis-invariant statements exclusively in terms of basis invariants.}
Rather, it is also possible to formulate them in terms of \textit{basis-covariant objects}.
These objects do transform under basis changes but, loosely speaking, 
their relative properties are basis invariant.

Perhaps surprising, formulating basis-invariant statement
in terms of basis-covariant objects sometimes leads to dramatic simplifications 
as compared to equivalent statements formulated in terms of basis invariants directly.
Below, focusing on $CP$ conservation in 2HDMs and 3HDMs, 
we will collect a few remarkable illustrations. In particular, this includes cases
where results in terms of basis invariants are not yet known due to their exceeding complexity. 
This gives convincing arguments that in sufficiently sophisticated models,
working with basis-covariant objects is the method of choice.

The paper is organized as follows.
In the next Section we will review the bilinear formalism, 
which is particularly useful in order to find basis-covariant in the NHDM.
Then we describe the problem of explicit $CP$ conservation 
and various possibilities which exist in the NHDM. 
Next, we describe how this issue was solved in the 2HDM and 3HDM,
and finally summarize our findings.

\section{Bilinear formalism}\label{section-bilinear}

\subsection{Bilinears}

Let us start with a brief review of the bilinear formalism of the 2HDM \cite{Nagel:2004sw,Ivanov:2005hg,Nishi:2006tg,Maniatis:2006fs,Maniatis:2007vn,Ivanov:2006yq,Ivanov:2007de,Nishi:2007dv}.
The most general renormalizable 2HDM Higgs potential constructed from two Higgs doublets
$\phi_a$, $a=1,2$ can be compactly written as
\be
V=Y_{ab} (\phi^{\dagger}_a\phi_b)+Z_{abcd} (\phi^{\dagger}_a\phi_b)(\phi^{\dagger}_c\phi_d)\,.\label{potential}
\ee
It depends on the Higgs fields via gauge-invariant combinations $\phi_a^\dagger \phi_b$,
which can be arranged into components of a real-valued bilinears
\be
r_0 = \phi^\dagger_a \phi_a\,,\quad r_i = \phi^\dagger_a (\sigma^i)_{ab} \phi_b\,,\quad i=1, 2, 3\,,
\label{bilinears-2HDM}
\ee
where $\sigma^i$ are the familiar Pauli matrices.
Each $(r_0,r_i)$ in (\ref{bilinears-2HDM}) is in one-to-one correspondence with an electroweak gauge
orbit in the space of doublets $\phi_a$. The map (\ref{bilinears-2HDM}) from doublets $\phi_a$
to $(r_0,r_i)$ does not cover the entire $1+3$-dimensional space but only the region defined
by inequalities $r_0 \ge 0$ and $r_0^2 - r_i^2 \ge 0$.
A basis change transformation $\phi_a \to \phi'_a=U_{ab} \phi_b$ with $U \in U(2)$ leaves $r_0$ 
invariant and induces an $SO(3)$ rotation of the vector $r_i$. 
Since the map $SU(2) \to SO(3)$ is surjective, any $SO(3)$ rotation in the bilinear space can be realized
as a basis change in the space of two doublets.

Beyond two Higgs doublets, the approach remains the same but the complexity of the problem skyrockets.
In the 3HDM \cite{Nishi:2006tg,Ivanov:2010ww,Maniatis:2014oza}, 
we define $1+8$ gauge-invariant bilinear combinations $(r_0, r_i)$:
\be
r_0 = {1\over\sqrt{3}}\phi^{\dagger}_a\phi_a\,,\quad r_i = \phi^{\dagger}_a (t^i)_{ab}\phi_b\,,\quad 
i=1,\dots,8\,,\quad a=1,2,3\,.\label{bilinears}
\ee
Here, $t_i = \lambda_i/2$ are the generators of the $SU(3)$ algebra satisfying
\be
[t_i,t_j] = i f_{ijk} t_k\,,\quad \{t_i,t_j\} = {1 \over 3}\delta_{ij}\id_{3} + d_{ijk} t_k\,,\label{tensors}
\ee
with the $SU(3)$ structure constants $f_{ijk}$ and the fully symmetric $SU(3)$ invariant tensor $d_{ijk}$.
Explicit expressions for the components of $r_i$ and a list of non-zero components
of the  $SU(3)$ invariant tensors are given in the appendix.
Group-theoretically, $r_0$ is an $SU(3)$ singlet while $r_i$ transforms in the adjoint representation of $SU(3)$.

Unlike in the 2HDM, where the bilinears only had to satisfy $r_0 \ge 0$ and $r_0^2 - r_i^2 \ge 0$,
in the 3HDM they must satisfy an additional constraint \cite{Ivanov:2010ww}:
\be
d_{ijk}r_ir_jr_k + {1\over 2\sqrt{3}}r_0(r_0^2 - 3 r_i^2) = 0\,. \label{ddd}
\ee
Under a basis change in the space of Higgs doublets, $\phi_a \to \phi'_a = U_{ab} \phi_b$ with $U \in SU(3)$,
$r_0$ is invariant while $r_i$ rotates as a vector of $SO(8)$.
However, not all $SO(8)$ rotations in the adjoint space can be obtained in this way;
they must in addition obey the constraint \eqref{ddd} and, therefore, conserve the contraction $d_{ijk}r_ir_jr_k$.

\subsection{Constructions in the adjoint space}

When passing from Higgs doublets to bilinears, the space we work in becomes more complicated
but the objects we study get simpler. 
The potential $V$ becomes a quadratic, rather than quartic function of variables,
\be
V = M_0 r_0 + M_i r_i + \Lambda_{0}r_0^2 + L_i r_0 r_i + \Lambda_{ij}r_i r_j\,.\label{V-bilinears}
\ee
The bilinear approach and the generic expression for $V$ above hold for any NHDM. All components
of the tensors $Y_{ab}$ and $Z_{abcd}$ in \eqref{potential} fill 
$M_0$, $\Lambda_0$, the entries of the real vectors $M$ and $L$ (lying in the adjoint space\footnote{%
Here, we distinguish between the adjoint space of arbitrary real vectors $x \in \RR^{N^2-1}$
transforming under the adjoint representation of $SU(N)$ and the orbit space 
which is spanned by those vectors $r \in \RR^{N^2-1}$ which can be constructed from the doublets
and satisfy \eqref{ddd} and the inequalities above.} $\RR^{N^2-1}$), as well as the $(N^2-1)\times (N^2-1)$ real symmetric matrix $\Lambda$.

In the 2HDM, any $SO(3)$ rotation can be induced by a basis change.
Therefore, the matrix $\Lambda$ can always be diagonalized and its eigenvectors
can always be aligned with the axes of the adjoint space $(x_1,x_2,x_3)$.
These eigenvectors as well as the vectors $M$, $L$ are covariant objects and transform
in the same way under basis changes.
Using $SO(3)$ invariant tensors $\delta_{ij}$ and $\epsilon_{ijk}$,
one can contract these vectors and obtain basis invariants.

For the 3HDM, the potential \eqref{V-bilinears} contains
two 8D vectors $M$ and $L$ and the $8 \times 8$ real symmetry matrix $\Lambda$.
The lack of complete $SO(8)$ rotational freedom 
implies that it is not guaranteed anymore 
that $\Lambda$ can be diagonalized by a Higgs-basis change.
Nevertheless, $\Lambda$ can always be expanded over its eigensystem,
and eigenvalues and eigenvectors can be found numerically.

The fact that $SU(3)$ basis changes do not offer the full $SO(8)$ rotational freedom 
in the adjoint space certainly feels like a major nuisance factor. 
However it also offers at our disposal two additional invariant tensors $f_{ijk}$
and $d_{ijk}$. One can use them to define $f$- and $d$-products of any pair
of vectors $a$ and $b$ from the adjoint space:
\be
F_i := f_{ijk} a_j b_k\,,\quad D_i := \sqrt{3} d_{ijk} a_j b_k\,.\label{fd-products}
\ee
These products respect group covariance: vectors $F$ and $D$ transform as the adjoint $SU(3)$ representations.
These new products are at the heart of the basis-invariant algorithms
for detection of various $CP$ symmetries in the 3HDM \cite{Nishi:2006tg,Ivanov:2018ime}.

\section{Explicit $CP$-conservation in the 2HDM}

\subsection{Different forms of $CP$ symmetry in the 2HDM}

Consider the scalar sector of the 2HDM and  
suppose that it explicitly conserves $CP$.
Does this statement unambiguously specify the model (up to basis changes)?
The answer is no. There exist several distinct forms of $CP$ symmetry,
which cannot be mapped one to another by any basis change.
Depending on what kind of $CP$ symmetry one imposes, one obtains
physically distinct models.
This fact is known since long ago \cite{Ecker:1987qp,book,Weinberg:1995mt,Grimus:1995zi};
its application to the 2HDM was discussed at length, for example, in \cite{Ferreira:2009wh, Ferreira:2010yh}.
Here, we briefly repeat the classification to set up the notation.

The general $CP$ transformation defined in \eqref{GCP} depends on the matrix $X$,
whose form is basis-dependent.
However any $CP$ transformation possesses a basis-invariant feature: its order, 
that is, how many times one must apply it
to obtain the identity transformation.
Starting from an arbitrary unitary $X$, one can bring it to a block-diagonal form 
\cite{Ecker:1987qp,Weinberg:1995mt},
which has on the diagonal either pure phase factors or $2\times 2$ matrices of the following type:
\be
\mmatrix{c_\alpha}{s_\alpha}{-s_\alpha}{c_\alpha}\quad
\mbox{as in Ref.~\cite{Ecker:1987qp},}\quad \mbox{or}\quad
\mmatrix{0}{e^{i\alpha}}{e^{-i\alpha}}{0}\quad
\mbox{as in Ref.~\cite{Weinberg:1995mt}.}\label{block}
\ee
Applying the $CP$ transformation twice results in a Higgs family transformation with matrix $XX^*$. 
If it happens that $XX^*=\id$, which takes place at $\alpha = 0$ or $\pi$, 
the $CP$ transformation is of order 2, 
which we will generically denote by CP2. If $XX^*\not =\id$ but $(XX^*)^k =\id$,
which requires $\alpha$ to be a multiple of $\pi/k$,
we get a $CP$ transformation of order $2k$ denoted as CP$2k$.
If no finite $k$ exists such that $(XX^*)^k =\id$, that is, if $\alpha/\pi$
is not a rational number, we say that the $CP$ transformation
is of infinite order, which we will denote as CP$\infty$.

Let us now list the options available in the 2HDM.
\begin{enumerate}
\item
{\bf CP2.}
For a CP2 transformation, there always exists a basis in which 
it takes the standard form $\phi_a \to \phi_a^*$, that is, $X = \id$.
This model is referred to as the $CP$-conserving 2HDM;
in the classification of \cite{Ferreira:2009wh, Ferreira:2010yh}
it was denoted as $\mathfrak{CP}1$.\footnote{%
Reference \cite{Ferreira:2009wh} introduced the notation,
also used in \cite{Ferreira:2010yh}, of $\mathfrak{CP}1$,
$\mathfrak{CP}2$, and $\mathfrak{CP}3$, for models obtained in the 2HDM
by a GCP symmetry where $X X^* = 1$, $X X^* = -1$, and neither 1 nor $-1$,
respectively.
We use here the fraktur symbol because our CPn refers to a symmetry
where $(CP)^n=1$. Thus, in the 2HDM,
CP2 $ \mathfrak{\mathrel{\widehat{=}} CP}1$,
CP4 $\mathrel{\widehat{=}}\mathfrak{CP}2$,
and CPn with $n > 4$ or CP$\infty$ $\mathrel{\widehat{=}} \mathfrak{CP}3$.}
In the adjoint space, the standard $CP$ transformation
corresponds to the mirror reflection: $x_{1,3} \to x_{1,3}$, $x_2 \to - x_2$.
In a different basis, this transformation is still a mirror reflection
but with respect to a different axis in the bilinear space.
Due to the full $SO(3)$ rotational freedom, any mirror reflection
with respect to an arbitrary direction in the bilinear space can be 
transformed via a basis change to the reflection with respect to $x_2$.

The necessary and sufficient condition for the 2HDM potential to possess a CP2 symmetry
is the existence of the \textit{real basis}, that is, a basis in which all coefficients are real \cite{Gunion:2005ja}.
The challenge is how to detect the existence of the real basis in a basis-invariant way.
Below, we will list two approaches to solve this problem.
\item
{\bf CP4.} This transformation implies $XX^*\not =\id$, but $(XX^*)^2=\id$,
which requires $\alpha = \pi/2$ in \eqref{block}.
This transformation has a remarkably simple geometric interpretation in the adjoint space:
it is the \textit{point reflection} $x_i \to - x_i$. 
In the classification of \cite{Ferreira:2009wh, Ferreira:2010yh} it was denoted as $\mathfrak{CP}2$.
The 2HDM incorporating $\mathfrak{CP}2$ was dubbed 
in \cite{Maniatis:2007de} the maximally $CP$-symmetric model.

This geometric picture clearly shows that imposing CP4 on the 2HDM scalar potential
is equivalent to simultaneously imposing three CP2s, each performing
a mirror reflection with respect to axes $x_1$, $x_2$, and $x_3$.
Thus, the maximally $CP$-symmetric model is certainly distinct from the usual $CP$-conserving 2HDM.
\item
{\bf Higher-order $CP$.}
If $\alpha$ in Eq.~\eqref{block} is not a multiple of $\pi/2$, there is no further simplification possible,
and the transformation manifests itself in the bilinear space as a generic rotary reflection (improper rotation).
It was dubbed as $\mathfrak{CP}3$ in the classification of \cite{Ferreira:2009wh, Ferreira:2010yh}.
Group-theoretically, one can define transformations of finite or infinite order,
but all of them have the same effect on the scalar potential of the 2HDM:
the potential will be invariant under a $O(2) \times \Z_2$ symmetry group
in the bilinear space.
\item
{\bf Combining two CP2s.}
Finally, one can construct a 2HDM by imposing two different CP2 symmetries at once.
Depending on the choice of these symmetry transformations and on their commutation properties,
one can end up with different models. 
In particular, one can arrive in this way at a model which cannot be
obtained just by imposing any single GCP.
For example, if one CP2 is the standard $CP$ (mirror reflection with respect to $x_2$)
and the other CP2 is based on $X = \mathrm{diag}(1, \, -1)$ (reflection with respect to $x_1$),
then the two GCPs commute, and the resulting model is known as the $\Z_2$-symmetric 2HDM
as it can be enforced by a single sign flip of one of the Higgs doublets.\footnote{%
Group-theoretically, imposing two different GCPs, in this case, is equivalent to imposing one GCP and a 
group of family symmetries. Thus, in this way one does not arrive at a completely new, previously overlooked 2HDM.}
\end{enumerate}

It turns out that for all of the cases listed above, 
the total symmetry group can always be factorized as $\mathrm{CP2}\times H$, where $H$ is a family symmetry
not involving any $CP$ transformation.
In other words, the more exotic $CP$ transformations of the cases 2--4 above are related to CP2
by adjoining it with elements from $H$, which act in the adjoint space as
\begin{enumerate}
\addtocounter{enumi}{1}
\item $180^\circ$ rotation in the $(x_1,x_3)$ plane;
\item rotation by any angle in the $(x_1,x_3)$ plane and $180^\circ$ rotation in the $(x_1,x_2)$ plane;
\item $180^\circ$ rotation in the $(x_1,x_2)$ plane.
\end{enumerate}
Consequently, for the 2HDM there are two ways to detect the enlarged symmetry groups: 
either by detecting the presence of multiple $CP$ symmetries (corresponding to the composition 
of usual $CP$ with unitary symmetries), or by directly detecting the unavoidable presence of the additional unitary symmetries.
Crucially, this does not apply for the 3HDM where there are genuinely distinct $CP$ symmetries that cannot be factored out, see Sec.\,\ref{sec:3hdm}.
Here, for the 2HDM, we focus on the first strategy and show how to detect the presence of multiple $CP$ symmetries.

\subsection{Explicit $CP$-conservation via $CP$-odd basis invariants}

All the different forms of $CP$ symmetry have a common consequence: 
all $CP$-odd physical observables are zero.
Thus, to detect physical $CP$ invariance with respect to \textit{any} form of $CP$ transformation,
one has to make sure that all $CP$-odd basis invariants are zero.

Constructing \textit{a} $CP$-odd basis invariant out of the couplings $Y_{ab}$ and $Z_{abcd}$
is a rather straightforward exercise \cite{Botella:1994cs}. The challenge is to find the minimal number
of $CP$-odd invariants such that setting them to zero implies
that \textit{all other} $CP$-odd invariants are zero as well.
This problem was first solved in 1994 for the 2HDM after electroweak symmetry breaking \cite{Lavoura:1994fv},
while the solution for the 2HDM before symmetry breaking was discovered in the mid-2000's
\cite{Branco:2005em,Davidson:2005cw,Gunion:2005ja}. 
Four $CP$-odd basis-invariant combinations were constructed, 
labeled $I_{Y3Z}$, $I_{2Y2Z}$, $I_{3Y3Z}$, $I_{6Z}$ in \cite{Gunion:2005ja},
according to the powers of tensors $Y_{ab}$ and $Z_{abcd}$ used.
Setting these invariants to zero ($I_i = 0$) implies that all other $CP$-odd invariants vanish, too.
This gives the necessary and sufficient
condition for the 2HDM scalar sector to be explicitly $CP$-conserving, which can be checked in any basis.
Very recently, a powerful method based on Hilbert series and plethystic logarithm 
was proposed in \cite{Trautner:2018ipq} which allows one to efficiently construct the full ring of $CP$-even and $CP$-odd invariants.
This offers a shortcut to find these four $CP$-odd basis invariants, and allows for a concise proof 
that the vanishing of these four invariants is indeed sufficient for explicit $CP$ conservation. 

$CP$-odd basis invariants can also be constructed using the bilinear formalism
\cite{Ivanov:2005hg,Nishi:2006tg,Maniatis:2007vn}.
Since $CP$-transformations always correspond to reflections in the adjoint space, 
$CP$-odd invariants can be constructed as triple products of the vectors constructed from
$L$, $M$, and $\Lambda$.
Defining \mbox{$L^{(p)}_i := (\Lambda^p)_{ij} L_j$} and \mbox{$M^{(p)}_i := (\Lambda^p)_{ij}M_j$},
and denoting the triple product as $(A,B,C)  := \epsilon_{ijk} A_i B_j C_k$,
one can construct the four $CP$-odd invariants
\be
{\cal I}_1 = (M, M^{(1)}, M^{(2)})\,, \quad 
{\cal I}_2 = (L, L^{(1)}, L^{(2)})\,, \quad 
{\cal I}_3 = (M, L, M^{(1)})\,, \quad 
{\cal I}_4 = (M, L, L^{(1)})\,.\label{triple-products}
\ee
Also, one can prove that the model is explicitly $CP$-conserving if and only if all four invariants ${\cal I}_i=0$.
The relation between these ${\cal I}_i$ and $I_{Y3Z}$, $I_{2Y2Z}$, $I_{3Y3Z}$, and $I_{6Z}$ of \cite{Gunion:2005ja} 
was established in \cite{Nishi:2006tg}.

The conditions $I_i = 0$ or  ${\cal I}_i=0$ indicate that the 2HDM possesses \textit{a} $CP$ symmetry.
However, one cannot distinguish whether it is just a single CP2,
or a higher-order $CP$, or a combination of several $CP$ symmetries imposed simultaneously.
These different forms of $CP$ invariance do lead to physically distinct $CP$-conserving 2HDMs.
But $CP$-odd invariants, by construction, cannot tell the difference between the physically distinct models
as they are not sensitive to the matrix $X$ in \eqref{GCP}.
Thus, it is mandatory to go beyond $CP$-odd invariants in order to recover this information.

\subsection{Explicit $CP$-conservation via $CP$-even basis invariants}

It is known that the presence of $CP$ violation can be detected exclusively
via $CP$-even invariants. For example, in the Standard Model,
precise quark sector measurements of $|V_{ud} V_{us}|$,
$|V_{cd} V_{cs}|$, and $|V_{td} V_{ts}|$ imply that
the unitarity triangle has a non-vanishing area, which is a measure of $CP$ violation.\footnote{%
We thank Jo\~ao Silva for reminding us of this example.}
What is more important is that $CP$-even invariants, being non-zero for $CP$ conserving models, 
can reveal which form of $CP$ symmetry is imposed.
The distinction appears in the form of additional relations among these invariants.

Let us illustrate this statement using the bilinear formalism. 
For the sake of the argument it suffices to treat the simplified case where $L = 0$ by assumption.
We are then left with one real 3D vector $M$ and the real symmetric $3\times 3$ matrix $\Lambda$.
In total, they have $6+3=9$ components.
The basis-change freedom is characterized in the bilinear space by the group $SO(3)$.
Thus, all inequivalent models can be characterized by $9-3=6$ basis-invariant parameters. 
A possible choice is\footnote{%
As stressed in \cite{Trautner:2018ipq}, the trace basis for invariants of $\Lambda$ 
may not be the most convenient choice for many applications. Nevertheless, it suffices for our argument here. }
\be
\Tr\Lambda\,,\quad \Tr\Lambda^2\,,\quad \Tr \Lambda^3\,,\quad
m_0 \equiv M_i M_i\,,\quad m_1 = M_i \Lambda_{ij} M_j\,,\quad m_2 = M_i (\Lambda^2)_{ij} M_j\,.
\label{collection}
\ee
All higher-order invariants are then expressible in terms of these six invariants, for example,
\bea\label{m3}
m_3 &=&
m_2 \Tr\Lambda - {1\over 2}m_1[(\Tr\Lambda)^2- \Tr\Lambda^2] + m_0 \det\Lambda\, ,
\\[1mm]\label{m4}
m_4 &=&
m_3 \Tr\Lambda - {1\over 2}m_2[(\Tr\Lambda)^2- \Tr\Lambda^2] + m_1 \det\Lambda\, .
\eea
Suppose, the model had a CP2 symmetry. Since we have set $L = 0$, the only remaining
non-trivial $CP$-odd invariant in Eq.~\eqref{triple-products} is 
\be
{\cal I} = I_{3Y3Z} = \epsilon_{ijk} M_i (\Lambda M)_j (\Lambda^2 M)_k\,. \label{I3Y3Z}
\ee
The condition ${\cal I} = 0$ is compact and basis invariant,
but it does not distinguish what particular CP2 we have imposed:
The usual $CP$, which amounts to $x_2 \to - x_2$,
a different mirror reflection $x_1 \to - x_1$, or both of them simultaneously.
One can also square ${\cal I}$ in order to express it exclusively via $CP$-even invariants,
\be
{\cal I}^2 = m_0 m_2 m_4 + 2 m_1 m_2 m_3 - m_0 m_3^2 - m_2^3 - m_4 m_1^2\,.\label{I2}
\ee
Using \eqref{m3} and \eqref{m4}, this can further be reduced to the invariants of Eq.~\eqref{collection}.
Setting this expression to zero represents the basis-invariant condition for explicit $CP$ conservation
written in terms of $CP$-even invariants.
However, this single relation is still equivalent to ${\cal I} = 0$ and cannot by itself distinguish
which, or how many reflections are imposed in addition.
We need an \textit{additional} relation among $CP$-even invariants to settle the issue.

To derive it, let us first fix the basis.
The intermediate relations will rely explicitly on this basis choice, 
but the final result will be basis-independent.\footnote{%
It is crucial that basis-independent relations are sought in the end.
The specific basis choice is just an auxiliary tool for constructing basis-independent relations.
In general, it is not recommended to start the classification of symmetries in a basis where $\Lambda$ is diagonal.
One may run into conditions which are \textit{not} renormalization group invariant. That is, 
they would not correspond to actual symmetry classes.
A specific example is given in Eqs.~(129)-(136) of \cite{Ferreira:2010yh}.
}
We choose the basis in which $\Lambda$ is diagonal, with eigenvalues $\lambda_{1,2,3}$,
which are, generically, non-zero and distinct, while $M$ has initially three non-zero components.
In this basis, invariance under the usual $CP$ ($\mathfrak{CP}1$) implies $M_2=0$,
which allows us to write the invariants $m_k$ as
\be
m_k = \lambda_1^k M_1^2 + \lambda_3^k M_3^2\,, \quad k = 0, 1, \dots.
\ee
From here we deduce an extra relation among invariants $m_k$: 
\be
CP (x_2 \to - x_2):\qquad m_2 - m_1(\lambda_1+\lambda_3) + m_0\lambda_1\lambda_3 = 0\,. \label{CP0}
\ee
Although the set of eigenvalues is basis-independent,
their \textit{ordering} depends on the basis choice,
and this relation explicitly distinguishes $\lambda_2$ from $\lambda_{1,3}$.

Imposing a different mirror reflection would produce the analogous relations, 
\bea
CP (x_1 \to - x_1):&&m_2 - m_1(\lambda_2+\lambda_3) + m_0\lambda_2\lambda_3 = 0\,.\label{CP1}\\
CP (x_3 \to - x_3):&&m_2 - m_1(\lambda_1+\lambda_2) + m_0\lambda_1\lambda_2 = 0\,.\label{CP3}
\eea
Thus, after fixing the basis, we do distinguish among different mirror-reflection symmetries.
As a cross check, to return to the basis-independent formulation of the same condition,
one can multiply all three expressions and set the product to zero. 
After some algebra we get
\be
\left[M_1 M_2 M_3 (\lambda_1 - \lambda_2)(\lambda_2 - \lambda_3)(\lambda_3 - \lambda_1)\right]^2= 0\,,
\ee
which is exactly ${\cal I}^2 = 0$.

Let us now return to the fixed basis and impose (\ref{CP0}) and (\ref{CP1}) simultaneously.
Assuming that the $\lambda_i$ are different, we first get two simplified basis-dependent relations,
$m_1 = m_0\lambda_3$ and $m_2 = m_1\lambda_3$, and in general $m_{k+1} = m_k \lambda_3$, 
from which we deduce a new basis-independent relation:
\be
m_1^2 = m_0m_2\,.\label{m1m2m0}
\ee
Thus, imposing condition \eqref{m1m2m0} not only guarantees that the model is $CP$-conserving
(which follows from direct substitution  in Eq.~\eqref{I2}, showing that ${\cal I}^2$ vanishes),
but also fixes the specific $CP$-symmetry to be the one of case 4: namely, simultaneously imposing two commuting mirror reflections. 
This relation also signals the presence of a unitary $\Z_2$ symmetry, 
distinct of $CP$, with a clear geometrical interpretation, see Sec.\,\ref{subsection-covariant-2HDM}. 
This demonstrates that relations between $CP$-even invariants can indicate the presence of symmetries other than $CP$.

If one imposes the three commuting CP2 symmetries simultaneously,
which in the 2HDM is equivalent to imposing CP4,
one notices, in the $\Lambda$-diagonal basis, that each component of the vector $M$ must be zero,
and therefore $m_0 = 0$. This is the $CP$-even basis-invariant condition for existence of the CP4 symmetry.

Notice that we needed to fix a basis here and perform the intermediate calculations in a
basis-dependent manner in order to arrive at a new relation among $CP$-even invariants.
This is not necessary in general and a universal approach to derive such relations, 
even in the absence of the simplifying assumption $L=0$, has recently been introduced in \cite{Trautner:2018ipq}.
Because the final relation occurs between basis-invariant quantities, and since it does
correspond to a legitimate symmetry, it is also renormalization group invariant. 
\enlargethispage{12pt}

\subsection{Explicit $CP$-conservation via basis-covariant objects}\label{subsection-covariant-2HDM}

The bilinear space approach outlined in section~\ref{section-bilinear}
offers a more direct insight into the structural properties of the scalar potential.
Due to the full $SO(3)$ rotational freedom in the adjoint space, 
the scalar sector of the 2HDM is fully specified by the eigenvalues of the matrix $\Lambda$
and by the orientation of the two real vectors $L$ and $M$ with respect to the 
eigenvectors of $\Lambda$. 
We stress that although all of these vectors are basis covariant objects,
their relative orientation can be specified in basis-invariant terms.
It is this relative orientation that encodes various forms of $CP$ symmetry in the 2HDM \cite{Ferreira:2010yh}.
\begin{enumerate}
\item
A model is explicitly $CP$-conserving if and only if there exists an eigenvector of $\Lambda$
which is orthogonal to both $M$ and $L$ \cite{Ivanov:2005hg,Nishi:2006tg,Maniatis:2007vn,Ferreira:2010yh}.
This geometric criterion is basis-invariant. 
Using linear algebra, one can rewrite this geometric observation 
in the form of four $CP$-odd invariants being equal to zero, cf.\,\eqref{triple-products}.
In fact, this is how these invariants were found in \cite{Nishi:2006tg}.
\item
Since CP4 implies the point reflection in the bilinear space, $x_i \to - x_i$,
imposing it on the 2HDM implies $M = 0$ and $L = 0$, with no restriction on $\Lambda$.
Each individual mirror reflection $x_1\to -x_1$, $x_2\to -x_2$, and $x_3\to -x_3$ along each eigenvector of $\Lambda$ is then a valid $CP$ symmetry.
\item 
Imposing symmetry under a higher-order $CP$ transformation implies that, in addition to $M = L = 0$, the matrix $\Lambda$ has a pair of degenerate eigenvalues.
This implies infinitely many mirror reflections as symmetries.
\item
Finally, imposing two commuting CP2 symmetries means 
that there exist \textit{two} eigenvectors of $\Lambda$ orthogonal to both $M$ and $L$.
Within the 3D adjoint space, this is equivalent to the statement
that there exists an eigenvector of $\Lambda$ \textit{parallel} to both $M$ and $L$.
The algebraic expression \eqref{m1m2m0} encodes precisely this information.
\end{enumerate}
There are two lessons that we learn from this list.
Firstly, formulating physically relevant features of the model solely in terms of basis invariants is not the only option.
Basis-invariant statements can also be expressed in terms of basis-covariant objects.
Secondly, and perhaps more importantly, we see that basis-invariant statement in terms of basis-covariant objects 
can often be formulated more concisely and derived more directly than via the brute force contraction of tensors. 
This applies to both, $CP$-even and $CP$-odd invariants.

Having seen the three approaches to establishing various forms of 
explicit $CP$ conservations in the 2HDM and having established a ``dictionary'' between some of them,
one may be tempted to think that it is just a matter of taste which approach to use.
However, in the next section we will consider the problem of explicit $CP$ conservation in the 3HDM and demonstrate
that working with basis-covariant objects substantially simplifies the analysis
as compared to working only with $CP$-even or $CP$-odd invariants.

\section{Explicit $CP$ conservation in the 3HDM}
\label{sec:3hdm}

Let us first remark that in the 3HDM, necessary and sufficient conditions 
for explicit $CP$ conservation in terms of basis invariants are not known.
Of course, one could follow the standard procedure 
and construct an arbitrary number of $CP$-odd invariants out of $Y_{ab}$ and $Z_{abcd}$ and require them to vanish,
\cite{Botella:1994cs,book,Branco:2005em,Davidson:2005cw,Gunion:2005ja,Varzielas:2016zjc}.
However, it is not known when one can stop this routine such that it is guaranteed that all higher-power 
$CP$-odd invariants also vanish. 
Also, constructing invariants and checking their algebraic independence is a cumbersome task, 
which is unavoidably delegated to a computer.
In addition, as in the 2HDM above, $CP$-odd invariants by themselves 
do not distinguish physically distinct forms of $CP$ symmetry. 
Therefore, one is again forced to involve $CP$-even invariants to distinguish different cases.
The method proposed in \cite{Trautner:2018ipq} may help overcome these difficulties, 
but it must first be extended to $SU(3)$.

It turns out that all of these conditions have been derived using basis-covariant objects, derived in the bilinear formalism.
The necessary and sufficient conditions for explicit CP2 conservation in the 3HDM were formulated already in 2006 in \cite{Nishi:2006tg}. 
The conditions for CP4 symmetry, as well as for simultaneously occurring CP2 and CP4 symmetries were derived recently in \cite{Ivanov:2018ime}. 
All these results are based on the (relative orientation between) vectors $M$ and $L$ and on the eigenvectors and eigenvalues 
of the $8 \times 8$ real symmetric matrix $\Lambda$. We will briefly recapitulate these results in this section.

\subsection{CP2 conservation in the 3HDM}

For any CP2 transformation, there exists a basis in which $X$ is the unit matrix
and the $CP$ transformation takes the standard form: $\phi_a \toCP \phi_a^*$, $a = 1,2,3$.
A necessary and sufficient condition for the potential \eqref{potential} to be explicitly CP2-conserving
is that in this basis all coupling coefficients are real.

In the adjoint space, in the basis where $X$ is the unit matrix, the CP2 transformation leaves invariant all vectors in the 5D subspace
$V_+ = (x_3,\, x_8,\, x_1,\, x_4,\, x_6)$,
while it flips the sign of all vectors in the 3D subspace $V_- = (x_2,\, x_5,\, x_7)$.
Therefore, the 3HDM potential is explicitly CP2-invariant if and only if there exists a basis 
in which the vectors $M, L \in V_+$ and $\Lambda$ is block-diagonal 
with a $5\times 5$ block in $V_+$ and a $3\times 3$ block in $V_-$.

The challenge then is to formulate this splitting in a basis-invariant form.
This problem was solved in \cite{Nishi:2006tg} with the aid of the eigenvectors of the matrix $\Lambda$.
First, consider the $3\times 3$ block of the matrix $\Lambda$ in the 3D subspace $V_-$.
Using the explicit expressions for the tensor $f_{ijk}$ given in \eqref{tensor-fijk}, 
one can verify that vectors of this subspace are closed under the $f$-product defined in \eqref{fd-products}: 
if $a, b \in V_-$, then $F_i = f_{ijk} a_j b_k \in V_-$.
Thus, CP2 invariance implies that there exist three mutually orthogonal eigenvectors of $\Lambda$,
which we denote $e$, $e'$, and $e''$, which are closed under the $f$-product.

Next, associating a vector $a$ in the adjoint space to a traceless Hermitian $3\times 3$ matrix $A$ 
\`{a} la $A := 2 a_i t_i$, 
one can recognize that the $f$-product of $a$ and $b$ corresponds to the commutator of
$A$ and $B$ \cite{Ivanov:2018ime}. Thus, CP2 invariance implies that the three traceless Hermitian matrices 
$E$, $E'$, and $E''$, corresponding to the three eigenvectors $e$, $e'$, and $e''$, 
are closed under taking commutators.
In short, $E$, $E'$, and $E''$ form a 3D subalgebra of $su(3)$.

There exist two options for 3D subalgebras of $su(3)$: $su(2)$ or $so(3)$.
In the adjoint space, the former corresponds, in a suitable basis, to the subspace $(x_1,x_2,x_3)$
while the latter corresponds to $V_-$. The difference between them is numerical:
$f_{123} = 1$ while $f_{257} = 1/2$. Therefore, if one finds that three orthonormal
eigenvectors of $\Lambda$ are closed under the $f$-product, in order to decide whether 
one has found the correct subalgebra, one computes the invariant 
\be
I_N = 2|f_{ijk} e_i e'_j e''_k|\,.\label{so3-invariant} 
\ee
If $I_N = 1$, we have found three eigenvectors which, in a suitable basis, span $V_-$. 
The last step is to check whether or not the vectors $M$ and $L$ have components in $V_-$.
If not, $M_i e_i=M_ie'_i=M_ie''_i = 0$ and $L_ie_i=L_ie'_i=L_ie''_i=0$.
If and only if all these conditions are satisfied, we have an explicitly CP2 conserving 3HDM \cite{Nishi:2006tg}. 

This sequence of checks represents the necessary and sufficient conditions for a CP2 symmetry in the 3HDM.
Despite involving basis-covariant objects, the ultimate conditions are basis-invariant and, therefore, can be checked in
any basis. It is clear that there must exist a formulation of these conditions in terms of basis invariants,
either $CP$-even or $CP$-odd. However, such a formulation is likely to be extremely complicated
as it has not been found yet. 

\subsection{CP4 conservation in the 3HDM}

Using three Higgs doublets, it is possible to construct a model whose only symmetry is
a $CP$-symmetry of order 4 (CP4) \cite{Ivanov:2011ae}.
This model, proposed in \cite{Ivanov:2015mwl} and denoted CP4 3HDM, has a peculiar property:
despite explicit $CP$ conservation, it contains irremovable complex coefficients in the scalar potential.
Unlike in the 2HDM, imposing a CP4 symmetry on the 3HDM does \textit{not} produce any accidental CP2 symmetry.
Thus, CP4 opens the path to a new model, physically distinct from any CP2-conserving situation and bearing its own interesting 
phenomenology \cite{Aranda:2016qmp,Ferreira:2017tvy,Ivanov:2017bdx,Haber:2018iwr,Cherchiglia:2019gll}. 
The basis-invariant necessary and sufficient conditions for the CP4 invariance in the 3HDM 
are not known in terms of basis invariants.
However, they are known in terms of basis-covariant objects \cite{Ivanov:2018ime}, and we will briefly
review them here. 

A $CP$ transformation of order 4 is a transformation $\phi_a \toCP X_{ab} \phi_b^*$,
whose matrix $X$, in an appropriate basis, takes the form
\be
X = \mmmatrix{0}{1}{0}{-1}{0}{0}{0}{0}{1}\,.
\label{X-CP4}
\ee
In this basis, CP4 acts on the adjoint space as
\bea
&& x_8 \to x_8\,,\quad (x_1, x_2, x_3) \to -(x_1, x_2, x_3)\nonumber\\
&&x_4 \to x_6\,, \quad x_6 \to -x_4\,, \quad x_5 \to -x_7\,, \quad x_7 \to x_5\,.
\eea
For the potential to be CP4 invariant, $M$ and $L$ must be aligned with $x_8$,
the only 1D subspace invariant under CP4, while $\Lambda$ must have the block diagonal form
\be
\Lambda =\mtrx{
    \fbox{\phantom{A}}_{\,3\times 3}& 0& 0\\
    0 & \fbox{\phantom{A}}_{\,4\times 4}& 0\\
    0 & 0 & \Lambda_{88}
    },
\label{Lamij-CP4-block}
\ee
with an arbitrary $3\times 3$ block in the subspace $(x_1,x_2,x_3)$ 
and very specific correlation patterns in the $4\times 4$ block
of the $(x_4,x_5,x_6,x_7)$ subspace, cf.\ \cite{Ivanov:2018ime}.

The block-diagonal form of $\Lambda$ in Eq.~\eqref{Lamij-CP4-block}
has two key features: the stand-alone direction $x_8$ and the 3D block in the subspace $(x_1,x_2,x_3)$.
Both features can be detected in a basis-invariant way.

First, one can take the $d$-product \eqref{fd-products} of any vector $a$ with itself:
$a_i \mapsto \sqrt{3}d_{ijk}a_j a_k$.
If the resulting vector is parallel to $a$, we call it \textit{self-aligned}.
If a vector is self-aligned, then there exists a basis in which it lies along the $x_8$ direction \cite{Ivanov:2018ime}.
Therefore, the basis-invariant criterion for splitting the $\Lambda_{88}$ entry from the rest 
is that there exists an eigenvector of $\Lambda$ which is self-aligned.
Let us denote this eigenvector as $e^{(8)}$.

Next, the $3\times 3$ block in the subspace $(x_1,x_2,x_3)$ implies that there are three eigenvectors
lying in this subspace.
Vectors in this subspace have a remarkable property:
they are $f$-orthogonal to the previously identified $e^{(8)}$.
The converse also holds: if there is a vector $a$ which is both orthogonal to $e^{(8)}$ 
(meaning $a_i e^{(8)}_i = 0$) and $f$-orthogonal to it 
(meaning the vector $f_{ijk} a_j e^{(8)}_k = 0$), 
then it must lie in the $(x_1,x_2,x_3)$ subspace and nowhere else.
Thus, we arrive at the basis-invariant condition for the $3\times 3$ block to split from the rest:
there must exist three mutually orthogonal eigenvectors of $\Lambda$
which are both orthogonal and $f$-orthogonal to the previously identified $e^{(8)}$.

Once the block-diagonal form of $\Lambda$ is established, 
what remains to be checked for the CP4 invariance is that all vectors are aligned with $e^{(8)}$. 
This concerns not only $M$ and $L$ but also vectors $K_i = d_{ijk}\Lambda_{jk}$
and $K^{(2)}_i = d_{ijk}(\Lambda^2)_{jk}$.
A byproduct of these checks is that the eigenvalues of $\Lambda$ in the $(x_4,x_5,x_6,x_7)$ subspace
are, at least, pairwise degenerate.
Having established the block-diagonal form of $\Lambda$
and having checked that $M$, $L$, $K$, and $K^{(2)}$ are aligned with $e^{(8)}$,
one concludes that the model respects a CP4 symmetry, see details in \cite{Ivanov:2018ime}.

\subsection{Combining CP4 and CP2}

From the classification of all discrete symmetry-based 3HDMs presented in \cite{Ivanov:2012ry, Ivanov:2012fp},
it follows that if CP4 is combined with any other symmetry, the resulting symmetry group
will unavoidably include a CP2 transformation. Thus, if one wishes to detect the presence of CP4-symmetry in the 3HDM
while at the same time excluding the presence of any other accidental symmetry, 
it is sufficient to check the \textit{absence} of any CP2 symmetry. 

In principle, one can check the conditions for CP2 independently from CP4.
However, using the algorithm explained above it turns out that it is actually shorter to check CP2 \textit{in addition} to CP4 \cite{Ivanov:2018ime}.
For this, one simply checks whether the $f$-product of eigenvectors of the previously identified $(x_4,x_5,x_6,x_7)$-subspace is itself an eigenvector of $\Lambda$.
If this is \textit{not} the case, one has found a pure CP4 3HDM model without any accidental symmetry. 

The problem of the basis-invariant recognition of an additional CP2 symmetry
in a CP4 symmetric 3HDM was also studied recently in \cite{Haber:2018iwr}.
Starting with a CP4 symmetric 3HDM, the authors discovered a 
$CP$-even basis-invariant, called $\mathcal{N}$, which is zero if and only if the model possesses an additional CP2 
symmetry that commutes with CP4. However, the constructed invariant is quite involved, being a high-degree polynomial of the quartic coefficients of the potential 
and the vacuum expectation values of the doublets.
In contrast, the algorithm presented in \cite{Ivanov:2018ime} is short, transparent, 
covers both commuting and non-commuting cases, and does not rely on vacuum expectation values.
This nicely illustrates the power of working with eigenvectors of $\Lambda$, i.e.\ basis covariant objects, 
as compared to manipulations directly with invariants. Since in the end one also wishes to link the
basis invariant statements to physical observables, as partly achieved in \cite{Haber:2018iwr}, it seems
likely that a combination of the different techniques could be fruitful.

\section{\label{sec:conclusions}Conclusions}

Physical observables must be independent of arbitrary choices of basis and, therefore, expressible
in terms of basis-invariant combinations of the Lagrangian parameters.
In addition, the potentially vast number of basis choices can obscure physically relevant features of a model. 
Together this justifies efforts to formulate various physically important features of New Physics models,
such as the issue of detecting explicit $CP$ conservation in multi-Higgs models, in terms of basis invariants.

However, in many cases it turns out to be prohibitively difficult to corner a physical feature of a model directly in terms of basis invariants.
The main message of the present paper is that in such cases it is still possible to establish basis-invariant statements as \textit{relations} between \textit{basis-covariant} objects.

Using relations between basis-covariant objects often offers a more 
direct and transparent way to basis-invariant statements.
This is firmly supported by the fact that a recently proposed systematic method for construction of basis invariants
uses basis-covariant objects as building blocks \cite{Trautner:2018ipq}. 
Resolving the \textit{substructure} of basis invariants in terms of \textit{basis-covariant} objects 
is essential there, also to derive relations between different basis invariants.
In the present paper, we have supported this message with many examples of conditions for $CP$ conservation
in the two- and three-Higgs-doublet models.

The 2HDM here serves as a warm-up exercise because the conditions are known
both in terms of invariants, and in term of basis-covariant objects. Therefore, a ``dictionary'' could be established.
We have shown that expressing various forms of $CP$ conservation via vectors in the adjoint
space, including the eigenvectors of the matrix $\Lambda$, leads directly to the basis-invariant conditions.

In the 3HDM, the power of using the (basis-covariant) eigenvectors of $\Lambda$ becomes evident. 
All cases of necessary and sufficient conditions for the 3HDM scalar potential
to be invariant under CP2, CP4, or CP2 and CP4 simultaneously, were derived in this approach.
A reformulation of these conditions directly in terms of basis invariants is not yet known.

All these examples support our point that, when dealing with sophisticated models 
with a large number of basis-choices, it can be much more efficient to work 
with basis-covariant building blocks rather than with basis invariants directly.
These building blocks do not have to be derived from the bilinear approach;
alternative paths are conceivable \cite{Trautner:2018ipq}.

Finally, we stress that the entire approach is in no way limited to 
our illustrative example of $CP$ violation in multi-Higgs models.
Analogous methods can be applied to achieve the basis-invariant detection of 
other symmetries in other models.

\subsection*{Acknowledgments}
We thank Jo\~{a}o~P.\ Silva for many useful discussions and numerous comments on the paper.
I.P.I.\ acknowledges funding from the Portuguese
\textit{Fun\-da\-\c{c}\~{a}o para a Ci\^{e}ncia e a Tecnologia} (FCT) through the FCT Investigator 
contract IF/00989/2014/CP1214/CT0004 under the IF2014 program,
and through the contracts UID/FIS/00777/2013, CERN/FIS-NUC/0010/2015, and PTDC/FIS-PAR/29436/2017,
which are partially funded through POCI (FEDER), COMPETE, QREN, and the EU.
I.P.I.\ also acknowledges the support from National Science Center, Poland, via the project Harmonia (UMO-2015/18/M/ST2/00518).
C.C.N.\ acknowledges partial support by Brazilian funding agencies Fapesp through grant 2014/19164-6 and CNPq through grant 308578/2016-3.
The work of A.T.\ is partly supported by a postdoc fellowship of the German Academic Exchange Service~(DAAD).
A.T.\ is grateful to Stuart Raby and the Physics Department of Ohio State University for hospitality during the completion of this work.
\appendix

\section{Invariant tensors of $SU(3)$}

In this appendix we give explicit expressions for the $SU(3)$ invariant tensors $f_{ijk}$ and $d_{ijk}$ 
of equation \eqref{tensors},
as well as for the $1+8$ gauge-invariant bilinear combinations $(r_0, r_i)$ defined in \eqref{bilinears}.
With the usual choice of basis for the Gell-Mann matrices $\lambda_i$, the totally anti-symmetric $f_{ijk}$ and symmetric $d_{ijk}$ have the non-zero components
\be
f_{123} = 1\,, \quad
f_{147} = -f_{156} = f_{246} = f_{257} = f_{345} = -f_{367} = {1\over 2}\,,\quad
f_{458} = f_{678} = {\sqrt{3} \over 2}\,,
\label{tensor-fijk}
\ee
as well as
\bea
&&d_{146} = d_{157} = - d_{247} = d_{256} = {1\over 2}\,,\qquad
\phantom{-} d_{344} = d_{355} = - d_{366} = - d_{377} = {1\over 2}\,,\nonumber\\
&& d_{118} = d_{228} = d_{338} = - d_{888} = {1\over \sqrt{3}}\,,\qquad
d_{448} = d_{558} = d_{668} = d_{778} = -{1\over 2\sqrt{3}}\,.\label{tensor-dijk}
\eea
The coefficient in the definition of the $SU(3)$ singlet $r_0$ is not fixed by the bilinear construction, 
but the exact normalization is inessential here. We use the definition adopted from \cite{Ivanov:2010ww}
but alternative normalization factors are possible, see e.g.\ \cite{Maniatis:2014oza}.
In the Gell-Mann basis, the bilinears $r_i$ read
\bea
&&r_1 + i r_2 = \phi_1^\dagger \phi_2\,,\quad
r_4 + i r_5 = \phi_1^\dagger \phi_3\,,\quad
r_6 + i r_7 = \phi_2^\dagger \phi_3\,,\nonumber\\
&&
r_3 = \fr{1}{2}(\phi_1^\dagger\phi_1 - \phi_2^\dagger\phi_2)\,,\quad
r_8 = \fr{1}{2\sqrt{3}}(\phi_1^\dagger\phi_1 + \phi_2^\dagger\phi_2 - 2 \phi_3^\dagger\phi_3)\,.\label{ri-explicit}
\eea


\begin{thebibliography}{99}

\bibitem{Lee:1973iz} 
  T.~D.~Lee,
  Phys.\ Rev.\ D {\bf 8}, 1226 (1973).
  doi:10.1103/PhysRevD.8.1226


\bibitem{Branco:2011iw} 
  G.~C.~Branco, P.~M.~Ferreira, L.~Lavoura, M.~N.~Rebelo, M.~Sher and J.~P.~Silva,
  Phys.\ Rept.\  {\bf 516}, 1 (2012)
  doi:10.1016/j.physrep.2012.02.002
  [arXiv:1106.0034 [hep-ph]].


\bibitem{Haber:2006ue} 
  H.~E.~Haber and D.~O'Neil,
  Phys.\ Rev.\ D {\bf 74}, 015018 (2006)
  Erratum: [Phys.\ Rev.\ D {\bf 74}, no. 5, 059905 (2006)]
  doi:10.1103/PhysRevD.74.015018, 10.1103/PhysRevD.74.059905
  [hep-ph/0602242].

\bibitem{book}
  G.~C.~Branco, L.~Lavoura and J.~P.~Silva,
  ``CP Violation,''
  Int.\ Ser.\ Monogr.\ Phys.\  {\bf 103}, 1 (1999).


\bibitem{Ecker:1981wv} 
  G.~Ecker, W.~Grimus and W.~Konetschny,
  Nucl.\ Phys.\ B {\bf 191}, 465 (1981).
  doi:10.1016/0550-3213(81)90309-6


\bibitem{Ecker:1983hz} 
  G.~Ecker, W.~Grimus and H.~Neufeld,
  Nucl.\ Phys.\ B {\bf 247}, 70 (1984).
  doi:10.1016/0550-3213(84)90373-0


\bibitem{Neufeld:1987wa} 
  H.~Neufeld, W.~Grimus and G.~Ecker,
  Int.\ J.\ Mod.\ Phys.\ A {\bf 3}, 603 (1988).
  doi:10.1142/S0217751X88000254


\bibitem{Ecker:1987qp} 
  G.~Ecker, W.~Grimus and H.~Neufeld,
  J.\ Phys.\ A {\bf 20}, L807 (1987).
  doi:10.1088/0305-4470/20/12/010


\bibitem{Grimus:1989qn} 
  W.~Grimus and H.~Neufeld,
  Phys.\ Lett.\ B {\bf 237}, 521 (1990).
  doi:10.1016/0370-2693(90)91218-Z


\bibitem{Botella:1994cs} 
  F.~J.~Botella and J.~P.~Silva,
  Phys.\ Rev.\ D {\bf 51}, 3870 (1995)
  doi:10.1103/PhysRevD.51.3870
  [hep-ph/9411288].


\bibitem{Branco:2005em} 
  G.~C.~Branco, M.~N.~Rebelo and J.~I.~Silva-Marcos,
  Phys.\ Lett.\ B {\bf 614}, 187 (2005)
  doi:10.1016/j.physletb.2005.03.075
  [hep-ph/0502118].


\bibitem{Davidson:2005cw} 
  S.~Davidson and H.~E.~Haber,
  Phys.\ Rev.\ D {\bf 72}, 035004 (2005)
  Erratum: [Phys.\ Rev.\ D {\bf 72}, 099902 (2005)]
  doi:10.1103/PhysRevD.72.099902, 10.1103/PhysRevD.72.035004
  [hep-ph/0504050].


\bibitem{Gunion:2005ja} 
  J.~F.~Gunion and H.~E.~Haber,
  Phys.\ Rev.\ D {\bf 72}, 095002 (2005)
  doi:10.1103/PhysRevD.72.095002
  [hep-ph/0506227].


\bibitem{Varzielas:2016zjc} 
  I.~de Medeiros Varzielas, S.~F.~King, C.~Luhn and T.~Neder,
  Phys.\ Rev.\ D {\bf 94}, no. 5, 056007 (2016)
  doi:10.1103/PhysRevD.94.056007
  [arXiv:1603.06942 [hep-ph]].

\bibitem{Trautner:2018ipq} 
  A.~Trautner,
  arXiv:1812.02614 [hep-ph].

\bibitem{Nagel:2004sw}
  F.~Nagel,
  ``New aspects of gauge-boson couplings and the Higgs sector'', PhD thesis (2004),
\url{http://archiv.ub.uni-heidelberg.de/volltextserver/4803/1/Dissertation_Nagel.pdf}

\bibitem{Ivanov:2005hg} 
  I.~P.~Ivanov,
  Phys.\ Lett.\ B {\bf 632}, 360 (2006)
  doi:10.1016/j.physletb.2005.10.015
  [hep-ph/0507132].


\bibitem{Nishi:2006tg} 
  C.~C.~Nishi,
  Phys.\ Rev.\ D {\bf 74}, 036003 (2006)
  Erratum: [Phys.\ Rev.\ D {\bf 76}, 119901 (2007)]
  doi:10.1103/PhysRevD.76.119901, 10.1103/PhysRevD.74.036003
  [hep-ph/0605153].


\bibitem{Maniatis:2006fs} 
  M.~Maniatis, A.~von Manteuffel, O.~Nachtmann and F.~Nagel,
  Eur.\ Phys.\ J.\ C {\bf 48}, 805 (2006)
  doi:10.1140/epjc/s10052-006-0016-6
  [hep-ph/0605184].


\bibitem{Maniatis:2007vn} 
  M.~Maniatis, A.~von Manteuffel and O.~Nachtmann,
  Eur.\ Phys.\ J.\ C {\bf 57}, 719 (2008)
  doi:10.1140/epjc/s10052-008-0712-5
  [arXiv:0707.3344 [hep-ph]].


\bibitem{Ivanov:2006yq} 
  I.~P.~Ivanov,
  Phys.\ Rev.\ D {\bf 75}, 035001 (2007)
  Erratum: [Phys.\ Rev.\ D {\bf 76}, 039902 (2007)]
  doi:10.1103/PhysRevD.76.039902, 10.1103/PhysRevD.75.035001
  [hep-ph/0609018].


\bibitem{Ivanov:2007de} 
  I.~P.~Ivanov,
  Phys.\ Rev.\ D {\bf 77}, 015017 (2008)
  doi:10.1103/PhysRevD.77.015017
  [arXiv:0710.3490 [hep-ph]].


\bibitem{Nishi:2007dv} 
  C.~C.~Nishi,
  Phys.\ Rev.\ D {\bf 77}, 055009 (2008)
  doi:10.1103/PhysRevD.77.055009
  [arXiv:0712.4260 [hep-ph]].


\bibitem{Ivanov:2010ww} 
  I.~P.~Ivanov and C.~C.~Nishi,
  Phys.\ Rev.\ D {\bf 82}, 015014 (2010)
  doi:10.1103/PhysRevD.82.015014
  [arXiv:1004.1799 [hep-th]].


\bibitem{Maniatis:2014oza} 
  M.~Maniatis and O.~Nachtmann,
  JHEP {\bf 1502}, 058 (2015)
  Erratum: [JHEP {\bf 1510}, 149 (2015)]
  doi:10.1007/JHEP10(2015)149, 10.1007/JHEP02(2015)058
  [arXiv:1408.6833 [hep-ph]].


\bibitem{Ivanov:2018ime} 
  I.~P.~Ivanov, C.~C.~Nishi, J.~P.~Silva and A.~Trautner,
  arXiv:1810.13396 [hep-ph].


\bibitem{Weinberg:1995mt} 
  S.~Weinberg,
  ``The Quantum theory of fields. Vol. 1: Foundations,''
      Cambridge University Press (1995).


\bibitem{Grimus:1995zi} 
  W.~Grimus and M.~N.~Rebelo,
  Phys.\ Rept.\  {\bf 281}, 239 (1997)
  doi:10.1016/S0370-1573(96)00030-0
  [hep-ph/9506272].


\bibitem{Ferreira:2009wh} 
  P.~M.~Ferreira, H.~E.~Haber and J.~P.~Silva,
  Phys.\ Rev.\ D {\bf 79}, 116004 (2009)
  doi:10.1103/PhysRevD.79.116004
  [arXiv:0902.1537 [hep-ph]].


\bibitem{Ferreira:2010yh} 
  P.~M.~Ferreira, H.~E.~Haber, M.~Maniatis, O.~Nachtmann and J.~P.~Silva,
  Int.\ J.\ Mod.\ Phys.\ A {\bf 26}, 769 (2011)
  doi:10.1142/S0217751X11051494
  [arXiv:1010.0935 [hep-ph]].


\bibitem{Maniatis:2007de} 
  M.~Maniatis, A.~von Manteuffel and O.~Nachtmann,
  Eur.\ Phys.\ J.\ C {\bf 57}, 739 (2008)
  doi:10.1140/epjc/s10052-008-0726-z
  [arXiv:0711.3760 [hep-ph]].


\bibitem{Lavoura:1994fv} 
  L.~Lavoura and J.~P.~Silva,
  Phys.\ Rev.\ D {\bf 50}, 4619 (1994)
  doi:10.1103/PhysRevD.50.4619
  [hep-ph/9404276].


\bibitem{Ivanov:2011ae} 
  I.~P.~Ivanov, V.~Keus and E.~Vdovin,
  J.\ Phys.\ A {\bf 45}, 215201 (2012)
  doi:10.1088/1751-8113/45/21/215201
  [arXiv:1112.1660 [math-ph]].


\bibitem{Ivanov:2015mwl} 
  I.~P.~Ivanov and J.~P.~Silva,
  Phys.\ Rev.\ D {\bf 93}, no. 9, 095014 (2016)
  doi:10.1103/PhysRevD.93.095014
  [arXiv:1512.09276 [hep-ph]].


\bibitem{Aranda:2016qmp} 
  A.~Aranda, I.~P.~Ivanov and E.~Jiménez,
  Phys.\ Rev.\ D {\bf 95}, no. 5, 055010 (2017)
  doi:10.1103/PhysRevD.95.055010
  [arXiv:1608.08922 [hep-ph]].


\bibitem{Ferreira:2017tvy} 
  P.~M.~Ferreira, I.~P.~Ivanov, E.~Jiménez, R.~Pasechnik and H.~Serôdio,
  JHEP {\bf 1801}, 065 (2018)
  doi:10.1007/JHEP01(2018)065
  [arXiv:1711.02042 [hep-ph]].

\bibitem{Ivanov:2017bdx} 
  I.~P.~Ivanov,
  JHEP {\bf 1802}, 025 (2018)
  doi:10.1007/JHEP02(2018)025
  [arXiv:1712.02101 [hep-ph]].

\bibitem{Haber:2018iwr} 
  H.~E.~Haber, O.~M.~Ogreid, P.~Osland and M.~N.~Rebelo,
  JHEP {\bf 1901}, 042 (2019)
  doi:10.1007/JHEP01(2019)042
  [arXiv:1808.08629 [hep-ph]].

\bibitem{Cherchiglia:2019gll} 
  A.~L.~Cherchiglia and C.~C.~Nishi,
  arXiv:1901.02024 [hep-ph].

\bibitem{Ivanov:2012ry} 
  I.~P.~Ivanov and E.~Vdovin,
  Phys.\ Rev.\ D {\bf 86}, 095030 (2012)
  doi:10.1103/PhysRevD.86.095030
  [arXiv:1206.7108 [hep-ph]].

\bibitem{Ivanov:2012fp} 
  I.~P.~Ivanov and E.~Vdovin,
  Eur.\ Phys.\ J.\ C {\bf 73}, no. 2, 2309 (2013)
  doi:10.1140/epjc/s10052-013-2309-x
  [arXiv:1210.6553 [hep-ph]].
  
  
\end{thebibliography}
\end{document}